\documentstyle[twocolumn,aps,psfig]{revtex}

\input{psfig}
\newcommand{\bold}[1]{\mbox{\boldmath $#1$}}    %       bold symbol
\newcommand{\pphi}{{\bold{\phi}}}               %       bold phi

\begin{document}

\title{Enhanced Event-by-Event Fluctuations in Pion Multiplicity\\
as a Signal of Disoriented Chiral Condensates at RHIC}

\author{M.~Bleicher, 
J.~Randrup\thanks{E-mail: Randrup@LBL.gov}, R.~Snellings, X.-N.~Wang}

\address{Nuclear Science Division, 
Lawrence Berkeley Laboratory\\
1 Cyclotron Road, Berkeley, California 94720, U.S.A.}

\date{June 20, 2000}

\maketitle
\begin{abstract}
The factorial moments of the pion multiplicity distributions 
are calculated with HIJING and UrQMD
and found to be independent of the $p_T$ range included,
in contrast to recent simulations with the linear $\sigma$ model
which leads to large enhancements for pions
with transverse kinetic energies below 200~MeV.
This %finding 
supports the use of
the ratio of the factorial moments of low and high $p_T$ pions 
as a signal of ``new'' physics at low momentum scales,
such as the formation of disoriented chiral condensates.
\end{abstract}

%\narrowtext

%%%%%%%%%%%%%%%%%%%%%%%%%%%%%%%%%%%%%%%%%%%%%%%%%%%%%%%%%%%%%%%%%%%%%%%%%%%%
A major goal of relativistic heavy-ion collision experiments
is to explore the phase diagram of hot and dense matter.
In addition to the anticipated transition
to the quark-gluon plasma phase\cite{QGP},
in which the individual hadrons have dissolved into a chromodynamic plasma
of quarks and gluons,
it is expected that chiral symmetry will be approximately restored
in the hot collision zone.
Signals of this latter type of phase transition
may arise from the subsequent non-equilibrium relaxation
of the chiral order parameter which is expected to exhibit
large-amplitude long-wavelength isospin-polarized oscillations
around the normal vacuum configuration,
often referred to as Disoriented Chiral Condensates \cite{DCC}.
One expected consequence would be an anomalous broadening in
the distribution of the neutral pion fraction.
However,
this observable poses serious practical challenges
and in fact early DCC experiments carried out at CERN and Fermilab
were not able to discern such a signal \cite{DCCexp}.
Therefore,
especially with the RHIC facility at Brookhaven National Laboratory
now becoming operational,
the need has intensified for theory to identify specific observables
that may be particularly informative.
In the present paper we address the pion multiplicity distribution. 

Recent simulations with the linear $\sigma$ model \cite{JR:rod}
(see below)
have suggested that the induced oscillations of the chiral order parameter
amplifies individual pion modes in the soft part of the spectrum and,
as a result, leads to enhanced fluctuations in the multiplicity distribution
of the produced pions. 
In that work, the multiplicity distributions were analyzed
in terms of their factorial moments,
\begin{equation}
f_m = \langle N(N-1)(N-2)\cdots (N-m+1)\rangle
\end{equation}
with $N$ being the particle multiplicity in the specified rapidity and momentum
region and $m$ denoting the order of the moment
(the average $\langle\cdot\rangle$ is taken over a sample of events).
We note that the first factorial moment is simply the mean multiplicity,
$f_1=\langle N\rangle$.
A more convenient normalization is provided
by the {\em reduced} factorial moments,
\begin{equation}
F_m = f_m/\langle N \rangle^m\quad ,
\end{equation}
since these are all unity if the multiplicity distribution is of Poisson form.
Thus, the deviation of the higher-order reduced factorial multiplicity moments
provide a direct indication that non-poissonian fluctuations are present. 
In Ref.~\cite{JR:rod} it was found that
the reduced factorial moments for soft pions
(those with a transverse kinetic energy below 200~MeV)
were significantly in excess of unity,
while those for the harder pions were remained consistent with unity.
This feature suggests that such an analysis be made for the early RHIC data.

However,
while the occurrence of reduced factorial moments in excess of unity
is an indication of non-trivial processes,
it is not necessarily a unique DCC signature.
Thus, it is important to ascertain how specific this phenomenon is.
For this purpose,
we compare the reported results of the linear $\sigma$ model
with those prediced by the two most commonly used event generators,
namely HIJING\cite{HIJING} and UrQMD\cite{UrQMD}.
HIJING is expected to give a particularly good approximation
to the particle production and fluctuations at high transverse momenta,
while UrQMD includes a detailed treatment of resonance formation and decays.
As both types of process lead to highly correlated pion production,
one may expect these models to yield some enhancement
of the multiplicity fluctuations.
Therefore it is of interest to perform a quantitative comparison
of the results. 

%--------------------------------------------------------------------------
Let us first briefly describe the calculations in Ref.~\cite{JR:rod}.
Using a semi-classical treatment of the linear $\sigma$ model \cite{JR:PRD55},
the chiral field 
$\pphi(\bold{r})=(\sigma(\bold{r}),\bold{\pi}(\bold{r}))$
was prepared to represent a rod-shaped source
with a bulk temperature of $T_0$ (typically 240~MeV)
and a radius of $R_0$ (6-10~fm).
The rod was then endowed with a longitudinal Bjorken scaling expansion
and the classical field equation was solved numerically
on a cartesian lattice,
using the comoving coordinates $(\tau,\eta)$ in place of $(t,z)$
(and starting from $\tau=\tau_0=1~{\rm fm}/c$),
\begin{equation}
[ {1\over\tau}\partial_\tau \tau\partial_\tau
-\partial_x^2-\partial_y^2-{1\over\tau^2}\partial_\eta^2
+\lambda(\phi^2-v^2)]\pphi=H\bold{e}_\sigma\ .
\end{equation}
Due to the imposed longitudinal expansion and, later on,
the self-induced transverse expansion,
the field amplitudes decrease rapidly.
When sufficient decoupling has been achieved,
a transverse Fourier resolution is made at each value of $\eta$.
From the resulting expansion coefficients of
$\bold{\pi}(\bold{r})$ and $\dot{\bold{\pi}}(\bold{r})$
one may then extract the coefficients
\begin{equation}
\bold{\chi}_{\bf k}(\eta)=\sqrt{\Omega_\perp}
\left[ \sqrt{m_k\over2} \bold{\pi}_{\bf k}(\eta)
+{i\over\sqrt{2m_k}}\dot{\bold{\pi}}_{\bf k}(\eta) \right]
\left({\tau\over\tau_0}\right)^{1\over2}
\end{equation}
which represent the probability amplitudes
for finding a particle with rapidity $\eta$
and transverse momentum $\bf k$.
(Here $\Omega_\perp$ is the cross section of the box
employed in the calculation,
$m_k$ is the transverse pion mass, $m_k^2=m_\pi^2+k^2$,
and the last factor compensates for longitudinal scaling expansion.)
Thus the expected number of pions within a certain rapidity interval
is given by $\bar{n}^{(j)}_{\bf k}=\int d\eta |\chi^{(j)}_{\bf k}(\eta)|^2$,
for each transverse momentum $\bf k$ and each charge state $j$.
(These quantities become constant after the decoupling has occurred.)
The actual multiplicity $n^{(j)}_{\bf k}$
was then selected from the associated Poisson distribution.
The total multiplicity $N$
(which is also Poisson distributed)
can be obtained subsequently by adding up all the pions
emitted within the specified phase space,
\begin{equation}
N^{(j)}_{\rm soft}=\sum_{k<k_0} n^{(j)}_{\bf k}\ ,\,\,\,\
N^{(j)}_{\rm hard}=\sum_{k>k_0} n^{(j)}_{\bf k}
\end{equation}
where $k_0$ denotes the maximum momentum of the soft pions.

In order to obtain sufficient statistics,
a number of independent Bjorken rods were treated.
Due to the thermal fluctuations,
each such ``event'' has a unique initial field configuration so, consequently,
the final states differ in detail.
In particular, the expected multiplicities $\bar{n}^{(j)}_{\bf k}$
fluctuate from one event to the next.
Additional fluctuation arises from the
subsequent sampling of the actual (integer) multiplicities $n$
based on the values expected for a particular event $\bar n$.
While the latter statistical process is poissonian by design,
the event-to-event fluctuation of $\bar n$ is generally not.
Depending on the isospin orientation of the relaxing chiral order parameter,
certain modes are amplified preferentially and this mechanism is the origin
of the anomalous fluctuations reflected in the factorial moments,
as discussed in Ref.~\cite{JR:rod}.

\begin{figure}[h]
\vskip 0mm
\vspace{1.0cm}
\centerline{\psfig{figure=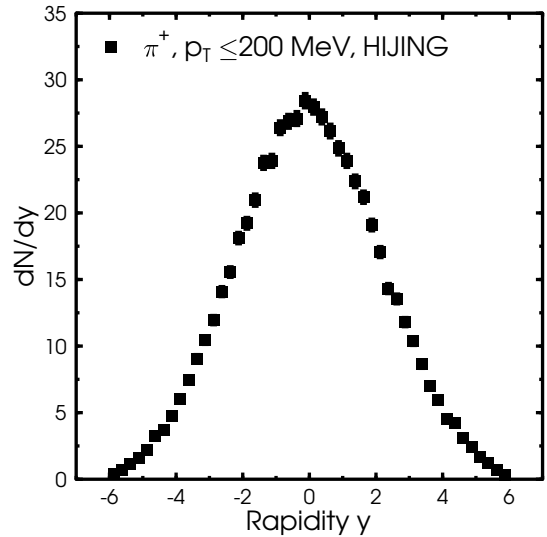,width=3.5in}}
\vskip 2mm
\caption{Rapidity distribution of soft positive pions
($p_T\leq~200~{\rm MeV}$), 
as generated by HIJING for a sample of central Au+Au events at RHIC
(impact parameter $b\leq3~{\rm fm}$).
The corresponding results for the negative and neutral pions are similar.
\label{dndy}}
\end{figure}

%--------------------------------------------------------------------------
In the present investigation,
we consider central Au+Au events (impact parameter $b\leq3~{\rm fm}$)
at the planned maximum beam energy at RHIC (100~GeV per nucleon).
In order to set the stage,
we show in Fig.~\ref{dndy} the average multiplicity of positive pions
with a transverse momentum below 200~MeV/$c$,
as obtained with the HIJING event generator \cite{HIJING}.
These soft pions constitute only a relatively small part
of the total number of pions
(the calculated mean $p_T$ of midrapidity pions
range from 330~MeV (UrQMD) to 370~MeV (HIJING)).

%..........................................................................
%..........................................................................

%..........................................................................
\begin{figure}[tbh]
\vskip 0mm
\vspace{-1.0cm}
\centerline{\psfig{figure=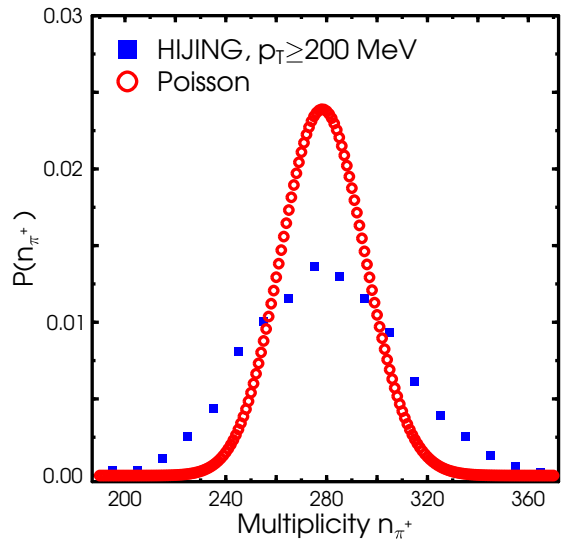,width=3.5in}}
\vskip 2mm
\caption{
The multiplicity distribution of positive pions in the mid-rapidity bin
($|y|<0.5$) for central Au+Au events at RHIC ($b\leq3~{\rm fm}$),
as generated by HIJING (solid squares).
The corresponding Poisson distribution having the same mean value
is also depicted (open circles).
\label{poisson}}
\end{figure}
%..........................................................................

We focus now on the multiplicity distribution in the mid-rapidity bin,
$|y|<0.5$.
Figure~\ref{poisson} shows the calculated multiplicity distribution
for positive pions of any energy,
as generated by HIJING for a sample of central Au+Au collisions.
Also shown is the Poisson distribution having the same mean multiplicity.
The calculated multiplicity distribution is significantly wider
than the Poisson distribution.
While this feature is due in part to the averaging over different
impact geometries,
it also reflects the presence of non-statistical components
in the microscopic processes.
The same feature still holds even if the HIJING analysis
is restricted to the soft pions only.
Moreover, similar features are also obtained with UrQMD
for either grouping of the pions (not shown in the figure).
Thus, it should be expected that the corresponding higher factorial moments
will exceed unity.

\begin{figure}[tbh]
\vskip 0mm
\centerline{\psfig{figure=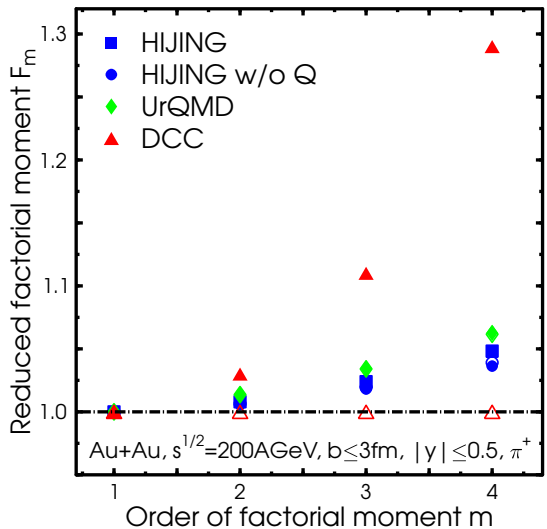,width=3.5in}}
\vskip 0mm
\vspace{-.0cm}
\caption{Reduced factorial moments $F_m$
for the multiplicity of positively charged pions
emitted at mid rapidity in central Au+Au collisions at the full RHIC energy,
as generated by HIJING with (squares) or without (circles) jet quenching
and by UrQMD (diamonds).
Also shown are the results obtained in Ref.~[4]	%\cite{JR:rod}
for idealized Bjorken rods (triangles).
Solid symbols represent soft pions
while open symbols represent hard pions.
}  
\label{fac1}
\end{figure}

That this is indeed borne out is evident from Fig.~\ref{fac1},
which shows the reduced factorial moments for both soft and hard pions
as calculated with various models:
HIJING with and without jet quenching,
UrQMD,
and the results obtained in Ref.~\cite{JR:rod}
for the Bjorken rods with the linear $\sigma$ model. 
It is evident that the latter results stand out:
Whereas all calculations with HIJING and UrQMD yield a rather similar
behavior, namely a gentle increase of $F_m$ with the order $m$,
the linear $\sigma$ model leads to reduced moments that remain close to unity
for the hard pions while increasing rapidly for the soft pions.

%..........................................................................
%..........................................................................

\begin{figure}[t]
\vskip 0mm
\centerline{\psfig{figure=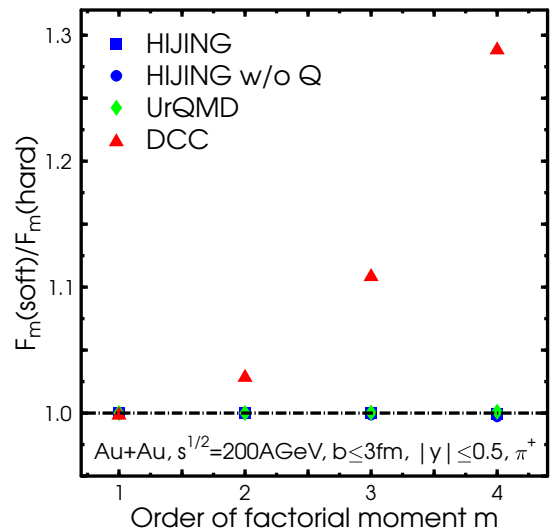,width=3.5in}}
\vskip 0mm
\vspace{-.0cm}
\caption{The ratios between corresponding values of the
calculated reduced factorial moments $F_m$ for soft and hard pions,
for the four cases displayed in Fig.~\ref{fac1}.
\label{fac2}}
\end{figure}

This qualitative difference in the behavior can be made more visible
by considering the ratios between the reduced moments for soft and hard pions,
as shown in Fig.~\ref{fac2}.
While all the HIJING and UrQMD calculations
predict a similar behavior
for pions with low and high transverse momenta,
the dynamical simulations with the linear $\sigma$ model
yield a strong enhancement of the fluctuations
in the number of low-$p_T$ pions.
In particular,
although both HIJING (with or without jet quenching) and UrQMD
produce some enhancement of the multiplicity fluctuations
(above a pure Poisson behavior),
neither one shows any distinction between soft and hard pions in this regard.
Thus it appears that a comparison of the factorial moments
of the soft and hard pion multiplicity distributions
may provide a useful observable which could indicate the presence
of interesting dynamics beyond what has been included in the standard
event generators. 
In particular, a relative enhancement of the low-$p_T$ factorial moments
may signal the formation of disoriented chiral condensates
and may therefore be used to gain experimental information 
on the global chiral properties.

%..........................................................................
%..........................................................................

%--------------------------------------------------------------------------

Let us finally emphasize some caveats associated with the present analysis.
Although the two Monte-Carlo models used here
successfully reproduce many features of particle production
in hadronic and nuclear collisions,
many issues regarding multiplicity fluctuations remain unresolved.
In particular,
the observed intermittency signal associated with very small rapidity bins
($\Delta y\ll1$) \cite{klm}  
cannot be reproduced satisfactorily with the models employed here.
However, for rapidity bins of the larger widths
employed in the present investigation ($\Delta y\approx 1$),
the data of Ref.~\cite{klm} can be reproduced
within the experimental error bars.
As for the linear $\sigma$ model,
it is expected that the semi-classical treatment underestimates the enhancement
(by a factor of two or more) \cite{JR:HIP}.
Moreover,
it applies only to baryon-free systems
(hence our focus is on mid-rapidity pions)
and current calculational capabilities
are limited to the relatively schematic Bjorken rod geometries
treated in Ref.~\cite{JR:rod}.
Therefore,
the results should be regarded as qualitative only.
Fortunately, though,
the conclusions presented here do not depend on the exact modeling
but rely only on the general features of the dynamics. 

%--------------------------------------------------------------------------
The present study was motivated by the recently reported finding that
dynamical simulations with the linear $\sigma$ model for idealized systems
lead to strong enhancements in the multiplicity fluctuations for soft pions,
while the harder ones display fluctuations of Poisson form.
Our present investigation has shown that
although both the HIJING and the UrQMD event generators
also produce multiplicity fluctuations in excess of pure Poisson statistics,
their magnitudes are relatively small (as compared to the DCC case)
and, importantly, they do not depend on the $p_T$ range considered.
These findings lend support to the adoption of this observable
as an indicator of DCC formation.

Thus, in conclusion, 
our analysis suggests that a strong relative enhancement
of the multiplicity fluctuations of pions with low transverse momenta
may provide a robust DCC signal in high-energy nuclear collision experiments,
such as those underway at RHIC. It is a special advantage that the basic
observable, namely the number of pions in a given $p_T$ range,
should be readily obtainable experimentally.
Thus, the suggested analysis qualifies as a Year-One task at RHIC.

%---------------------------------------------------------------------------
\section*{Acknowledgements}
This work was supported by the Director, Office of Energy Research,
Office of High Energy and Nuclear Physics,
Nuclear Physics Division of the U.S. Department of Energy,
under Contract DE-AC03-76SF00098
and it used resources of the
National Energy Research Scientific Computing Center at LBNL (NERSC).
One of us (M.B.) acknowledges support by the Alexander von Humboldt Foundation
as a Feodor Lynen Fellow.

%---------------------------------------------------------------------------

%---------------------------------------------------------------------------

\vfill
%---------------------------------------------------------------------------
{\small \noindent
{\em LBNL-46102: \hfill Physical Review C (Rapid Communication)}
}
\end{document}